%% file: main.tex
\documentclass[sigconf]{acmart}

\usepackage{enumitem}

\newcommand{\miniskip}{\vspace*{-.5\baselineskip}}

\newcommand{\todo}[1]{\textcolor{red}{#1}}

\copyrightyear{2020}
\acmYear{2020}
\setcopyright{acmcopyright}
\acmConference[ICTIR '20]{Proceedings of the 2020 ACM SIGIR International Conference on the Theory of Information Retrieval}{September 14--17, 2020}{Virtual Event, Norway}
\acmBooktitle{Proceedings of the 2020 ACM SIGIR International Conference on the Theory of Information Retrieval (ICTIR '20), September 14--17, 2020, Virtual Event, Norway}
\acmPrice{15.00}
\acmDOI{10.1145/3409256.3409834}
\acmISBN{978-1-4503-8067-6/20/09}

\begin{document}

\title[Bias in Conversational Search]{Bias in Conversational Search: \\ The Double-Edged Sword of the
  Personalized Knowledge Graph}

\author{Emma J. Gerritse}
\affiliation{\institution{ Radboud University}} 
		\email{emma.gerritse@ru.nl}
\author{Faegheh Hasibi}
\affiliation{\institution{ Radboud University}} 
\email{f.hasibi@cs.ru.nl}

\author{Arjen P. de Vries}
\affiliation{\institution{ Radboud University}} 
\email{a.devries@cs.ru.nl}


\begin{abstract}
	
Conversational AI systems are being used in personal devices, providing users with highly personalized content. Personalized knowledge graphs (PKGs) are one of the recently proposed methods to store users' information in a structured form and tailor answers to their liking. Personalization, however, is prone to amplifying bias and contributing to the echo-chamber phenomenon. In this paper, we discuss different types of biases in conversational search systems, with the emphasis on the biases that are related to PKGs. We review existing definitions of bias in the literature: people bias, algorithm bias, and a combination of the two, and further propose different strategies for tackling these biases for conversational search systems. Finally, we discuss methods for measuring bias and evaluating user satisfaction.


\end{abstract}



\maketitle

\input{introduction.tex}
\input{biastypes.tex}

\input{solution.tex}

\input{evaluation.tex}

\input{conclusion.tex}

\bibliographystyle{ACM-Reference-Format}
\bibliography{literature.bib}

\end{document}

%% file: introduction.tex
\section{Introduction}

Conversational search has gained great attention recently, driven by the success of conversational AI systems such as Alexa, Google Home, and their likes. These systems are being installed on personal devices such as watches, phones, and laptops, collecting more and more personal information from users. Ideas have been proposed to make use of this personal information and build personalized systems, in order to gain improved user satisfaction. One of the proposed approaches is building \emph{personalized knowledge graphs (PKGs)}, which is a structured form of information about ``entities personally related to a user, their attributes and the relations between them''~\cite{Balog:2019:Personal}. PKGs provide the system with (locally stored) rich information about the user, to tailor the answers as much to their liking as possible \cite{Balog:2019:Personal}. However, introducing such a data resource and personalizing search results is not without risks: personalization can lead to echo-chamber effect and reinforce bias~\cite{Bozdag:2013:BAF}.

Bias in search and recommendation systems can influence human decision making, contribute to societal and political biases, and  impact the health of our society~\cite{Olteanu:2019:FACTS}. Examples of bias in current search and social media websites include Google's anti-Brexit bias\footnote{\url{https://www.dailymail.co.uk/news/article-7605265/Google-facing-claims-anti-Brexit-bias-web-searches.html}}, Facebook's filter bubbles  during US presidential election\footnote{\url{https://www.theguardian.com/us-news/2016/nov/16/facebook-bias-bubble-us-election-conservative-liberal-news-feed}}, and Linkedin's gender bias\footnote{\url{https://time.com/4484530/linkedin-gender-bias-search/}}. 
Bias can get even more severe in a conversational setting because: (i) answers in conversational systems are concise and there is less chance of showing diversified results compared to the traditional ten blue links interface, and (ii) users tend to access information that are in line with their prior views~\cite{Koutra:2015:ECI}, and conversational systems provide personalized results to optimize for high user satisfaction and engagement, thereby intensifying bias. This raises several questions about bias in conversational search. 
Can forms of bias be introduced in this personalized search setting, which would not happen so easily in other search settings? 
How will conversational systems influence the behavior of their users? And, can users influence the behavior of their conversational agents? 
What are the sources of bias and how can we measure bias? How to mitigate bias while keeping users engaged and satisfied?

In this position paper, we discuss these questions from multiple angles. We consider the risk of different types of biases, and the places in the system architecture where these might be introduced; a major challenge being the difficulty to distinguish between \emph{bias} and \emph{preference}. We discuss how bias  can be introduced by search algorithms, specifically when constructing PKGs from users' personal data and using them to provide personalized responses. We further explain how PKGs play the role of a double-edged sword to on one hand amplify bias but on the other hand detect and mitigate bias (Section~\ref{sec:biastypes}).


Next, we argue that the community has to consider conversational search systems in regard to questions of bias, and find solutions to address it (Section~\ref{sec:solution}).  The main usage of a conversational system is a convenient form of tackling search; not an assistant to educate the user. If we would correct system results for bias too strongly, the debiasing process can only degrade user satisfaction. Conversational search therefore faces quite a challenge with respect to design ethics: not only do we need to think through the biases that we encounter and find solutions to counter those biases, we also have to do so in a manner that maintains user satisfaction. We outline possible approaches to help avoid the reinforcement of biases, while keeping the attraction of personalization to improve user satisfaction. Finally, we discuss methods to measure bias and user satisfaction (with respect to bias measures) in conversational systems (Section~\ref{sec:evaluation}).


%% file: biastypes.tex
\section{Types of bias in conversational search}
\label{sec:biastypes}
\citet{Baeza:2018:Bias} enumerates different types of bias on the web and groups them into three categories: (i) bias that involves only algorithms, (ii) bias that originates from people, and (iii) bias that involves both algorithm and people. We discuss the different types of biases we expect to encounter in conversational search along the same categories.


\miniskip
\subsection{Bias from Algorithms}
\label{ssec:biasa}
\paragraph{\textbf{Knowledge graph construction bias}} 
Knowledge graphs (KGs) are structured repositories of data and powerful means to provide a machine understandable form of knowledge. It is envisaged that public, domain-specific, and personalized knowledge graphs (PKGs) can be used to empower conversational search systems~\cite{Balog:2019:Personal}.  While generation of general purpose knowledge graphs (e.g., YAGO and WikiData) and domain-specific knowledge graphs (e.g., GeoNames and MusicBrainz) can involve bias, construction of personalized knowledge graphs can be even a greater source of bias.

PKGs can be built from social media feeds, search history \cite{He:2017:Measuring}, conversation history, and other sources of information for which users give the system access permission (e.g., online shops and contacts). The choice of data source by itself can already introduce bias in PKGs. If the system relies on publicly available social media feeds (e.g., Twitter~\cite{Yen:2019:Personal}), PKGs may be biased towards only one aspect of users' preferences: people often use different social media websites for different purposes; e.g., Twitter for work and Instagram for casual usage. Even if the system gets access to the content of a user's private social media feeds, the question remains whether the PKG is diversified enough or not: the user might avoid sharing some aspects of her life in social media. Therefore, conversational systems need to account for the incompleteness of PKGs and incorporate it into their search and recommendation algorithms. Otherwise, this can negatively affect users' satisfaction, as the user encounter undesired bias in search and recommendation results (e.g., observing only work-related events instead of diversified ones).   

Another source of bias in constructing PKGs is \emph{time}~\cite{Balog:2019:Personal}. Time-based events can greatly influence search behavior. For example, during election times people are more likely to share their opinion about politics, even when normally politics is not an important aspect of their lives. Another example is that during the Covid-19 outbreak in 2020, many people used social media to discuss the epidemiology, even though epidemiology is normally not in their interests. When using social media for knowledge graph construction, this can lead to some kind of time-based bias or time-based clutter, which will ``somehow'' have to be cleaned up after the event.

%
%

\paragraph{\textbf{Search algorithms bias}}
Bias can also occur in search and recommendation algorithms.
While not much research covers the biases in conversational search, some research has been carried out to measure bias in neural ranking models and embedding algorithms. 
In a recent study, \citet{Rekabsaz:2020:Neural} found that gender bias in document ranking is intensified when using neural models and in particular the ones that are based on contextual embeddings like BERT \cite{Devlin:2018:Bert}. 

Bias has been also observed in the word and graph embeddings themselves; e.g., social bias in WikiData graph embeddings~\cite{Fisher:2019:Measuring} and gender bias in word embeddings (like Word2Vec)~\cite{Bolukbasi:2016:Debiasing}. While debiasing embeddings may appear as an immediate solution to mitigate bias, incorporating debiased embeddings in information retrieval algorithms may not immediately affect the ranking outcome, as confirmed by some initial results in \cite{Gerritse:2020:Effect}. 

\miniskip
\subsection{Bias from People}
\label{ssec:biasp}
People with their activities and preferences themselves form another source of bias in conversational systems.
A study on age demographic of conversational systems' users in the United States showed that the most active users are 30-44 years old \footnote{https://voicebot.ai/2019/06/21/voice-assistant-demographic-data-young-consumers-more-likely-to-own-smart-speakers-while-over-60-bias-toward-alexa-and-siri/}. Indeed, people's preferences vary by their age group, and therefore a large population of a certain age group can introduce certain biases in the system. For example, a study on recommender systems~\cite{Zhao:2014:Demographic} shows that that preference of phone color differs per age group, and recommendations can be tailored to the age group of users. Not only age group, but also gender, location, and language of people can affect the outcome of conversational systems, often causing results biased towards the majority groups. 


\miniskip
\subsection{Bias from Algorithms and People}
\label{ssec:biasaap}
Conversational systems are highly interactive and are often designed to be personalized. These two features make conversational systems prone to a vicious cycle of bias that is generated by a biased user interacting with a biased algorithm, thereby creating a filter-bubble effect.  Personalized knowledge bases are double-edged swords in such a setting: they can be used to accelerate personalization and therefore providing even more biased results to the user, or they can be used to detect a biased user and help to diversify results. Drawing a line between personalization and diversification is a challenge and varies for different scenarios. Here, we enumerate three different scenarios as examples: 
\begin{itemize}[leftmargin=7mm]
    \item A user wants to order the same guitar strings (s)he previously ordered. However, these might not be the highest quality strings. Should the system recommend other strings or guide the user to buy the previously bought strings?
    \item A user has a certain political preference and asks a political question about her favorite party. Should the system takes users' preferences into account, or provide a neutral answer?
    \item A user searches for a known conspiracy, e.g., fake news about Covid-19 outbreak. Knowing that the user trusts the source of information based on available information in PKG, how the system should respond to the query?
\end{itemize}

In the following section, we propose some solutions that a system can take to handle bias.

%% file: solution.tex
\section{Possible strategies}
\label{sec:solution}

Conversational search systems can take a wide range of strategies against bias, from being completely ignorant to chastising users. In this section, we discuss three main strategies that these systems can take and argue against being ignorant.

\subsection{Ignorant}
In this setting, the system does not detect bias and takes no action against it. Depending on the search algorithm, the system may provide nonsensical responses or get along with users and provide them with what they ask. Such systems can fall into the trap of creating a vicious cycle of second-order bias (generated from people and algorithmic bias)~\cite{Baeza:2018:Bias} and then become their own enemy. An example of such a scenario is Microsoft's  AI chatbot that had to be taken down 16 hours after release, because of learning harmful intents from some Twitter users\footnote{\url{https://www.bbc.com/news/technology-35890188}}. We argue that being ignorant and \emph{not accounting for bias} is not an option for conversational search systems, especially because they are meant to work in a highly personalized setting.


\subsection{User in Control}
\begin{figure}[tb]
	\centering
	\includegraphics[width=\linewidth]{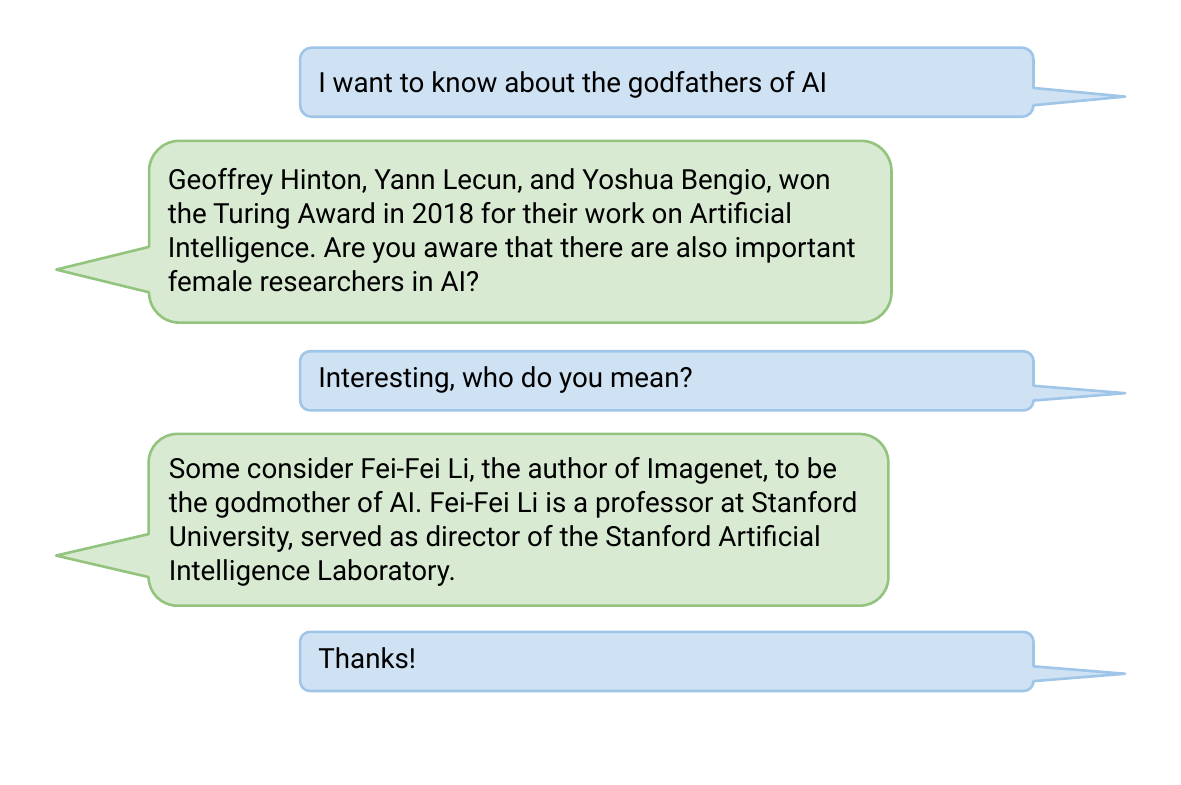}
	\caption{Example of the \emph{User in Control} strategy. The system detects the possible bias in the word 'godfathers', and offers the user also some interesting female answers.}
	\label{fig:con1}
\end{figure}

In this strategy, systems detect bias and offer the user control through its UI/UX, from being viciously biased or offering diversified results; see Figure~\ref{fig:con1} for an example. There is a wide spectrum of actions that systems take in this strategy, ranging from complete neutralization of results to playing along with users until reaching a certain threshold (e.g., inappropriate user requests). Choosing a spot in this wide spectrum of approaches is a challenge and requires carrying out user studies. In addition, users may have different tolerance for interacting with diversified answers: some may appreciate it and others may abandon the system altogether. Most of the current commercial systems choose to ``play along'' with users or take a middle approach of not answering or providing ``Do not know'' answers to gender and sexual questions~\cite{Cercas:2018:metoo}. This, however, may not be an ideal setting for the well being of all users. Providing users with an option to choose a degree of diversification (and therefore mitigate bias), e.g.\ with a UI control like a slider, may alleviate these issues. This does however raise additional implementation challenges, especially for data-driven approaches (like neural networks). Incorporating knowledge graphs and in particular PKGs can help detecting bias in the first place and further control it by providing factual and diversified responses (e.g., alternating between nodes of PKGs) to reduce bias.


\begin{figure}[tb]
	\centering
	\includegraphics[width=\linewidth]{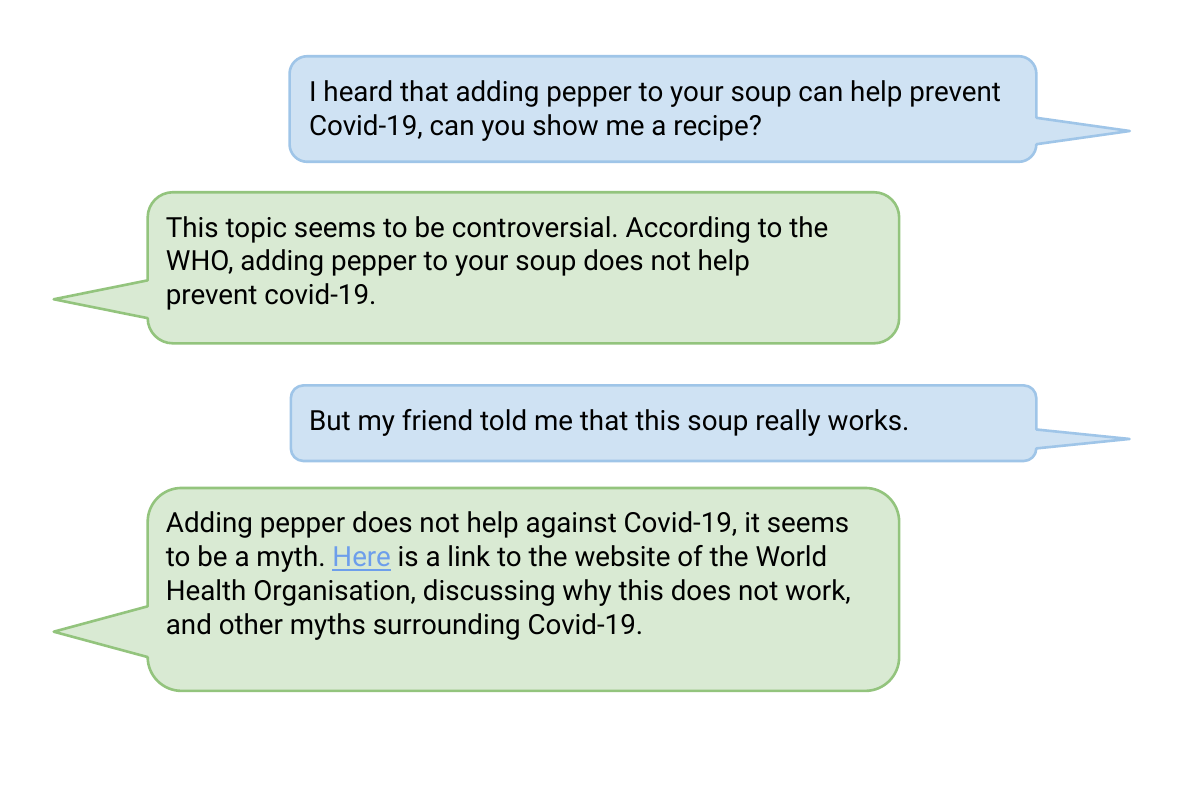}
	\caption{Example of the \emph{Reflective} strategy, where the system detects that the user is biased towards fake news, and the system supplies them with an alternative.}
	\label{fig:con2}
\end{figure}


\subsection{Reflective}
Another approach is being reflective about bias: both user and system detect bias and explicitly reflect it to each other. So here debiasing is the responsibility of both user and system. 

In the case of user bias, the system notifies the user in an appropriate language and suggests alternatives (see Figure \ref{fig:con2}). Educating users with an explicit response may be considered as an over-reaction, however, it is desired for some cases; e.g., health-related fake news in a pandemic like COVID-19, sexual harassment, or cases when the user is gathering confrontational viewpoints for an argument. 

Users may also correct an unnecessary bias of system and help to create a balance between bias and preference. Examples of such scenarios include: users changing their interests and opinions over time, or searching for something either on the spur of the moment or on someone else's request. 
In such cases, systems should provide the user with the source of bias and the option to mediate it. PKGs, while being a source of bias, can help the explainability of the system and enable users to influence the behavior of systems by editing their own PKG.

%% file: evaluation.tex
\section{Evaluation}
\label{sec:evaluation}
In this section, we propose methods to measure bias in conversational search and further evaluate user satisfaction when applying debiasing methods. Here, we do not aim to provide formal evaluation measures for bias, focusing instead on methods that can identify bias.

\subsection{Measuring bias}
\label{ssec:measuringb}
Researchers have raised challenges of measuring bias in online information~\cite{Pitoura:2018:OMB} and search engines~\cite{Mowshowitz:2005:MSE}. Measuring bias in conversational search raises even more challenges, due to the difficulties of evaluating the systems themselves. While test collections like TREC CAST~\cite{Dalton:2019:TREC} and QuAC~\cite{Choi:2018:Quac} exist, they are often restricted to the question-answering part of conversational systems. Evaluating multiple aspects of a conversational system is often performed with user experiments, and is mainly performed by considering general user satisfaction. We argue that a system's ability to handle bias should be measured separately and be  an integral element of user evaluations in personalized conversational search systems.

Another option to measure bias in conversational systems is to have an evaluation method that does not require relevance judgments. In \cite{Rekabsaz:2020:Neural}, one such method is discussed for computing gender bias in search results. This is performed by computing a term frequency based score with predefined gendered words. Similar approaches can be employed to measure the bias of an answer to a query, not only with respect to gender but other aspects like political bias.


It is important to note that bias is a relative concept and depending on the context, a biased answer may not be always undesired. 
One way of measuring bias is computing the results with and without PKGs; e.g., by using the general purpose knowledge graphs.  If the difference is rather large, then it indicates that personalization has occurred in a large degree (which is not always an undesired behavior).
This is indeed under the assumption that global knowledge graphs are fair representation of information. One can also generate the results with an \emph{opposite knowledge graph}, where nodes and relations of user's PKG are replaced with opposite information. We acknowledge that this concept does not generalize, as opposite information may not be available for all nodes. However, this solution may provide interesting insights for certain biases like gender and politics. Consider for example swapping ``Female'', ``right-wing'', and ``conservative'' with ``Male'', ``left-wing'', and ``progressive'', and compare the bias scores as described in~\cite{Rekabsaz:2020:Neural}. This indicates the influence of these nodes on the results. If there is a large difference between the two polarities, it means that personalization has occurred to a large extent. 

\subsection{Measuring user-satisfaction}
\label{ssec:measuringu}
 
Changing system behavior to handle bias, either by diversifying results or offering a reflective option, may change user satisfaction. It is expected that users do not appreciate the  feeling of being judged or educated by their conversational assistant. It is therefore wise to first study the best strategy of handling bias, before investigating technical aspects of offering these solutions.  To this end, users satisfaction can be measured by wizard experiments, similar to~\cite{Vtyurina:2017:Exploring}. Here, a wizard is a human who pretends to be a conversational search system and interacts with a human. These studies can be performed by instructing a wizard to offer perspectives to questions of human test subjects and measure user satisfaction.

%% file: conclusion.tex
\section{Conclusion}
\label{sec:conclusion}
In this paper, we touched upon some aspects of bias in conversational search. We raised questions about bias in the personalized search setting, especially using personalized knowledge graphs. We discussed the challenge of distinguishing between bias and preference, caused by the inherent bias of user preferences and also limitations of offering diversified results in conversational settings.  We enumerated different types of bias present in conversational search, being algorithmic bias, people bias and bias introduced by both algorithm and people. We also suggested possible solutions for handling bias, based on diversification and mutual reflection of bias between user and system. Lastly, we suggested methods of measuring bias and evaluating user satisfaction. Bias in conversational search is an aspect that requires extra attention and we urge the community to keep it in mind while developing conversational search  systems.